\documentclass{elsart}

 \usepackage{graphicx}
\usepackage{amssymb,amsmath}
\usepackage{bm}% bold math
\usepackage{dcolumn}% Align table columns on decimal point
\usepackage{soul} 

\usepackage{color}
\usepackage{ulem}
\begin{document}

\begin{frontmatter}



%\title{Spin glass transition of anisotropic $XY$ model}
\title{Electromechanical properties of ferroelectric polymers: Finsler geometry modeling and a Monte Carlo study}

\author{V. Egorov$^{1}$, O. Maksimova$^{1}$, H. Koibuchi$^{2}$, C. Bernard$^{3,4}$,}
\author{ J-M. Chenal$^{5}$, O. Lame$^{5}$, G. Diguet$^{3}$, G. Sebald$^{3,5}$,}
\author{ J-Y. Cavaille$^{3,5}$ and T. Takagi$^{6}$ }
%\ead{koibuchi@mech.ibaraki-ct.ac.jp}

\address{                    
  $^{1}$ Cherepovets State University,  Cherepovets,  Russian Federation \\
  $^{2}$ National Institute of Technology (KOSEN), Sendai College,  Natori, Japan \\
  $^{3}$ ELyTMaX UMI 3757, CNRS-University de Lyon-Tohoku University International Joint Unit, Tohoku University, Sendai, Japan \\
 $^{4}$ Frontier Research Institute for Interdisciplinary Sciences (FRIS), Tohoku University, Sendai, Japan \\
  $^{5}$ Materials Engineering and Science (MATEIS), CNRS, INSA Lyon UMR 5510, Universit$\acute{\rm e}$ de Lyon,  Villeurbanne Cedex, France \\
  $^{6}$ Tohoku Forum for Creativity,  Tohoku University, Sendai, Japan
}

\begin{abstract}
Polyvinylidene difluoride (PVDF) is a ferroelectric polymer characterized by negative strain along the direction of the  applied electric field.  However, the electromechanical response mechanism of PVDF remains unclear due to the complexity of the hierarchical structure across the length scales. As described in this letter, we employ the Finsler geometry model as a new solution to the aforementioned problem and demonstrate that the deformations observed  through Monte Carlo simulations on 3D tetrahedral lattices are nearly identical to those of real PVDF. Specifically, the simulated mechanical deformation and polarization are similar to those observed experimentally. 
\end{abstract}

\end{frontmatter}

% main text
%\section{}
%\label{}

%%%%%%%%%%%%%%%%%%%%%%%%%%%%%%%
\section{Introduction}
%%%%%%%%%%%%%%%%%%%%%%%%%%%%%%%
Ferroelectric polymers such as polyvinylidene difluoride (PVDF) exhibit strongly coupled mechanical and electrical properties. Because of their large deformations under external electric fields, PVDF and its copolymers are commonly used to design actuators \cite{bae2019pvdf,lee2019pvdf,burnham2017strain}. In contrast to typical classical ferroelectrics, PVDF exhibits an unusual negative longitudinal piezoelectric coefficient, i.e., the bulk thickness decreases in the direction of the applied electric field. Moreover, the mechanical strain is proportional to the square of the polarization \cite{furukawa1990electrostriction}, which implies that the field-induced deformation in the PVDF is a result of the electrostriction. However, the strain and polarization are linearly correlated under low intensity electric fields \cite{furukawa1990electrostriction}. This complex behavior occurs because both piezoelectricity and electrostriction are associated with scales ranging from the monomer to polymeric domain sizes, and this complex multiscale behavior can be reflected through the ``electromechanical properties'' of PVDF.

Broadhurst et al. \cite{broadhurst1978piezoelectricity} demonstrated that the piezoelectricity can be attributed to a mechanism known as the dimensional effect. According to this mechanism, the dipoles are rigid, and the polarization behavior can be attributed to the macroscopic shape deformation. Katsouras et al. \cite{katsouras2016negative} indicated that the piezoelectric nature of PVDF is a result of two microscopic mechanisms: changes in the lattice constant and coupling of the crystalline and amorphous parts.

Several studies based on first principles and molecular dynamics simulations have been devoted to PVDF-based polymers \cite{su2004density,bystrov2013molecular,dong2016first,satyanarayana2012molecular}. However, purely atomistic approaches cannot be used to effectively examine polymers because polymer chains consist of thousands of monomers that cannot be processed through these calculations. Nevertheless, well-developed phenomenological models for ferroelectrics exist, which can reproduce the polarization-electric field (PE) curves \cite{ducharme2000intrinsic}. In addition, several researchers combined molecular dynamics and phenomenological approaches \cite{ahluwalia2008multiscale}. Despite the notable research conducted in this domain, the molecular mechanisms of the piezoelectric effect and electrostriction in PVDF remain unclear.

In Finsler geometry (FG) modeling, the anisotropy of mechanical, optical or other properties is represented through a Finsler metric that effectively deforms the discrete elastic energy  \cite{koibuchi2014monte}. Consequently, the corresponding interaction among the molecules is direction dependent and hence anisotropic. The FG modeling technique has been successfully applied to evaluate the deformation of materials with mechanical anisotropy, such as rubbers and soft biological materials \cite{takano2017j} as well as the shape transformation of liquid crystal elastomers under external electric fields \cite{proutorov2018finsler}. In this letter, we extend the FG model to describe the unusual piezoelectric effect pertaining to ferroelectric polymers by defining the Finsler metric using an internal degree of freedom $\sigma$ corresponding to the polarization \cite{Egorov_2019}.
%%%%%%%%%%%%%%%%%%%%%%%%%%%%%%%
\section{Model \label{model}}
%%%%%%%%%%%%%%%%%%%%%%%%%%%%%%%
Consider the continuous form of the Hamiltonian for the tensile energy,
\begin{eqnarray}
\label{continuous-eneg}
S_1=\int \sqrt g  g^{ab} \frac {\partial \vec r} {\partial x^a} \cdot\frac {\partial \vec r} {\partial x^b} d^3 x, 
\end{eqnarray}
where $\vec r (\in {\bf R}^3)$ is a position vector of the material and $x^a (a \!=\! 1, 2, 3)$ denotes the local coordinates. The symbol $g^{ab}$ indicates the inverse of the Finsler metric $g_{ab}$, as described in the following text, and $g$ is the corresponding determinant. If $g_{ab}$ is the Euclidean metric, the expression (\ref{continuous-eneg}) represents the classical Gaussian bond potential, which is a 3D extension of the linear chain model \cite{doi2013soft}.

\begin{figure}[h!]
\centering
\includegraphics[width=12.5cm]{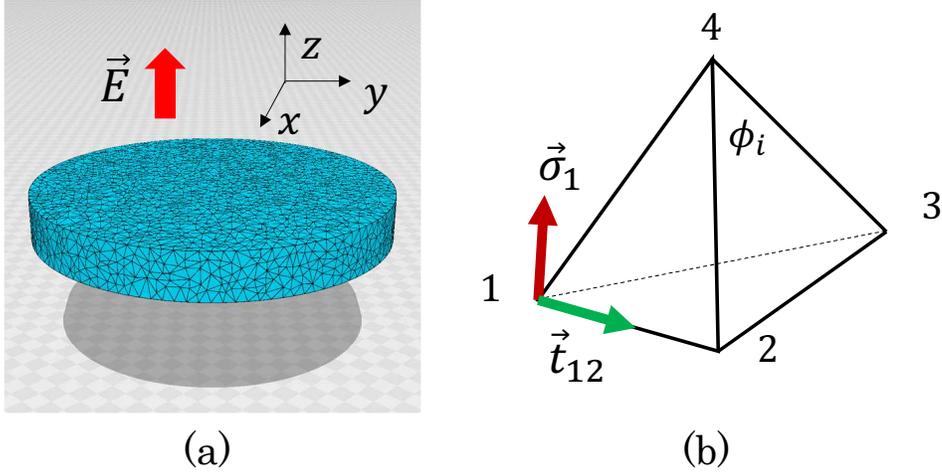}
\caption{\label{fig-1}(a) Thin cylinder discretized by tetrahedrons of size $N\!=\!10346$ and used for the simulations. (b) Tetrahedron on which the Finsler metric is defined.}
\end{figure}

To define the Finsler metric, we consider a 3D thin cylinder discretized by tetrahedrons by using Voronoi tessellation (Fig. \ref{fig-1}(a)). The ratio of the cylinder height to its diameter is 0.125, and the lattice is the same as that used in Ref.  \cite{proutorov2018finsler}.
A variable ${\vec \sigma_i} ~(\in S^2: {\rm unit \; sphere})$ is introduced for the polarization at vertex $i$ of a tetrahedron (Fig. \ref{fig-1}(b)). In contrast to the case of a liquid crystal elastomer, as described in Ref.  \cite{proutorov2018finsler}, ${\vec \sigma_i}$ is assumed to be a polar variable. 
Using ${\vec {\sigma_i}}$, we define the unit Finsler length along bond $ij$ such that
\begin{eqnarray}
\label{finsler-length}
v_{ij}=\sqrt{1-(\vec\sigma_i \cdot \vec t_{ij})^2}+v_0,
\end{eqnarray}
where $\vec t_{ij}$ is the unit tangential vector from vertex $i$ to vertex $j$. 
The expression presented as Eq. (\ref{finsler-length}) is new and different from that used in \cite{proutorov2018finsler}. 
In accordance with this new definition,  the tetrahedrons shrink along the direction of ${\vec \sigma}$. 
The parameter $v_0$ serves to strengthen/weaken the anisotropy in the mechanical properties, and it is assumed that $v_0\!=\!0.1$. At vertex 1 of tetrahedron 1234 (Fig. \ref{fig-1}(b)), the metric  is defined as
\begin{eqnarray}
\label{Finsler-metrc}
g_{ab}= 
\begin{pmatrix}
v_{12}^{-2} & 0 & 0 \\
0 & v_{13}^{-2} & 0 \\
0 & 0 & v_{14}^{-2}
 \end{pmatrix}.
\end{eqnarray}
%%%%%%%%%%%%%%%%%%%%%%%%%%%%%%%%%%%%%%%%%%%%%
%%%%%%%%%%%%%%%%%%%%%%%%%%%%%%%%%%%%%%%%%%%%%
Using the equivalence of the expressions with respect to the substitutions of indices $(1\rightarrow2, 2\rightarrow3, 3\rightarrow4, 4\rightarrow1)$, $ (1\rightarrow3, 2\rightarrow4, 3\rightarrow1, 4\rightarrow2)$,
$\cdots$ and by replacing the differentials with differences, we obtain
\begin{eqnarray}
\label{discrete-energy}
\begin{gathered}
S_1=\sum_{ij}\Gamma_{ij} l_{ij}^2,\, \Gamma_{ij}=\frac{1}{4\bar N}\sum_{\rm tet}\gamma_{ij}({\rm tet}), \\
\gamma_{12}=\frac{v_{12}}{v_{13}v_{14}}+\frac {v_{21}}{v_{23}v_{24}} , \gamma_{13}=\frac{v_{13}}{v_{12}v_{14}}+\frac {v_{31}}{v_{32}v_{34}} ,\\
\gamma_{14}=\frac{v_{14}}{v_{12}v_{13}}+\frac {v_{41}}{v_{42}v_{43}}, \gamma_{23}=\frac{v_{23}}{v_{21}v_{24}}+\frac {v_{32}}{v_{31}v_{34}},\\
\gamma_{24}=\frac{v_{24}}{v_{21}v_{23}}+\frac {v_{42}}{v_{41}v_{43}}, \gamma_{34}=\frac{v_{34}}{v_{31}v_{32}}+\frac {v_{43}}{v_{41}v_{42}},
\end{gathered}
\end{eqnarray}
where $\bar N$ is the mean total number of tetrahedrons sharing bond $ij$ and is given by $\bar N\!\simeq\!4.84$ for the considered lattice. 

We intuitively demonstrate the deformation mechanism of the shape of the liquid crystal elastomer by the external electric field.   
Note that $\Gamma_{ij}$ in $S_1$ functions as the local tension coefficient of the bond $ij$. If the variable $\sigma$ is aligned through an external electric field, owing to the interaction implemented in $v_{ij}$ in Eq. (\ref{finsler-length}), almost all $\Gamma_{ij}$ along the direction of the electric field become larger than those along the perpendicular direction. Consequently, the corresponding bond lengths $\ell_{ij}$  along the field direction decrease. These aspects pertain to an intuitive explanation of the effect of the interaction of the polarization vector $\sigma$ and polymer position ${\vec r}$ implemented in $v_{ij}$. 

The full Hamiltonian of the model including certain additional terms can be expressed as
\begin{eqnarray}
\begin{gathered}
\label{Hamiltonian}
S=\lambda S_0 +\gamma S_1+\kappa S_2 + S_3 + \alpha S_4,  \quad (\gamma=1)\\
S_0=-\sum_{(i,j)} \vec\sigma_i\cdot\vec\sigma_j,\quad S_2=\sum_{i} \left[1 -\cos(\phi_i-\pi/3)\right],\\ 
S_3=-\sum_i \vec\sigma_i \cdot \vec E,\quad S_4=-\sum_i (\vec\sigma_i \cdot \vec E)^2.
\end{gathered}
\end{eqnarray}
The unit of energy is $k_BT$, which is fixed as $k_BT\!=\!1$ in the simulations, where $k_B$ and $T$ denote the Boltzmann constant and temperature, respectively. Notably, the simulation unit is defined by the relation $k_BT\!=\!1$ for the energy, and the relation for the lattice spacing $a\!=\!1$ pertains to the length. The lattice spacing $a$ is introduced such that all the quantities with the unit of length are multiplied by $a$.  
Moreover, the coefficient $\gamma$ of $S_1$ is fixed as $\gamma\!=\!1$ for simplicity. This condition can be ensured by rescaling $T$ in the Boltzmann factor $\exp (-S/k_BT$) such that $\gamma /T\!=\!1/T^\prime$. In accordance with the new $T^\prime$,  all other coefficients are rescaled. The symbol $T$ denotes the temperature instead of $T^\prime$,  and the expression for $S$ in Eq. (\ref{Hamiltonian}) can be obtained. 
$S_2$ corresponds to the strength against the shear and bending deformations, and $\phi_i$ denotes the internal angle of the triangle at vertex $i$ (Fig. \ref{fig-1}(b)). The parameter $\kappa$ denotes the stiffness, and it influences the resistance against all deformations of the tetrahedron except the similarity deformation. 
As described in the subsequent sections, $\kappa$ effectively determines the strength of the electromechanical coupling; i.e., this parameter is macroscopically reflected in the slope of the strain-polarization curve.

$S_0$ represents the interaction between two nearest neighbors $\sigma_i$ and $\sigma_j$. Such a Heisenberg spin Hamiltonian is widely used in simulations for ferroelectric domain structures \cite{gerasimov2018study,novakovic2013pseudo,wang1994ferroelectric}. In these models, the energy $\langle \vec\sigma_i\cdot\vec\sigma_j \rangle\!=\!-S_0/N$ without an external electric field increases with increasing $\lambda$. This phenomenon is the same as that considered in our FG model except $\sigma$ interacts with the lattice deformation through $v_{ij}$ in Eq. (\ref{Finsler-metrc}). Thus, the remnant polarization is controlled by $\lambda$.

To describe the interaction between the polarization and external electric field $\vec E$, we introduce $S_3$ and $S_4$. Note that the coefficient for $S_3$ is assumed to be 1 for simplicity. This condition can be ensured by rescaling the electric field from $E$ to $E^\prime$ such that $aS_3(E)\!+\!bS_4(E)\!=\!S_3(E^\prime)\!+\!\alpha S_4(E^\prime)$ for any combination of $a(\not=\!0)$ and $b$. Using the relations $E^\prime\!=\!a E$ and $\alpha\!=\!b/a^2$, we can obtain the expression $S_3\!+\!\alpha S_4$ in Eq. (\ref{Hamiltonian}).  The symbol $E$ denotes the electric field instead of $E^\prime$. 
$S_3$ represents a classical PE field interaction, and $S_4$ is the corresponding quadratic extension. In this case, the shape of the PE curve is expected to be dependent on the parameter $\alpha$. For small $\alpha$ values, the PE hysteresis loop is squarelike, which is typical of crystalline materials. The loop becomes more rounded with increasing $\alpha$. Notably, $\lambda$ and $\alpha$ determine the polarization strength of the model, and the electromechanical coupling can be controlled by varying the parameters $\kappa$ and $v_0$.

The partition function $Z$ is defined as 
\begin{eqnarray}
\label{part-func}
Z=\sum_\sigma \int \prod_{i=1}^{N-1}d{\vec r}_i \exp(-S),
\end{eqnarray}where $\sum_\sigma$ denotes the sum over all possible configurations of $\sigma$, and $\int \prod_{i=1}^{N-1}d{\vec r}_i$ denotes $3(N-1)$-dimensional integration, where the center of mass is fixed at the origin of ${\bf R}^3$.  

%%%%%%%%%%%%%%%%%%%%%%%%%%%%%%%
\section{Simulation results}
%%%%%%%%%%%%%%%%%%%%%%%%%%%%%%%

The Monte Carlo (MC) simulation technique is used to study the behavior of a 3D cylinder under an external field $\vec E$ along the $z$ axis (Fig. \ref{fig-1}(a)). The vertex position ${\bf r}$ and polarization $\vec \sigma$ are updated with the Metropolis algorithm \cite{metropolis1953equation}. To suppress meaningless configurations, we apply several constraints on ${\bf r}$. Specifically, the volume of a tetrahedron cannot be negative and exceed tens of the initial mean value. The squared bond lengths must be less than $10\ell_0^2$, where $\ell_0$ is the mean initial bond length. To ensure that the thin cylinder is horizontal, a hard-wall potential is introduced such that the $z$ component of ${\bf r}$ lies in $(-H_0, H_0)$, where $H_0$ is the mean initial height of the cylinder in the equilibrium configuration for $E\!=\! 0$ and is evaluated through test simulations. These constraints do not influence the equilibrium configurations, and this aspect can be verified by the relation $S_1/N\!=\!3(N-1)/(2N)\!\simeq\!1.5$, the plot of which is presented below. This relation can be attributed to the scale-invariant property of the partition function \cite{proutorov2018finsler} and is used to verify whether the simulation program and simulations are sufficiently correct. 

In this letter, we compare our simulation results with the experimental data of uniaxially drawn PVDF measured at a low electric field frequency of 0.1 Hz \cite{furukawa1990electrostriction}. In the experimental data, the polarization and strain vs. electric field form hysteresis loops. However, the standard MC technique allows the examination of only the equilibrium properties. Therefore, we simulate only parts of the return loop in which the polarization aligns along the electric field and switching is not expected, as in the equilibrium configuration. The data are acquired at frequencies less than 1 kHz \cite{mai2015field}, and hence, a value of 0.1 Hz is considered to be sufficiently low for equilibration.

In the simulations, we set $\lambda\!=\!0.345$ to ensure that the system is in the ferroelectric phase. $\lambda\!=\!0.345$ is selected owing to the following: According to the test simulations, the polarization $\langle\sigma_z\rangle$ starts to saturate at $\langle\sigma_z^{\rm max}\rangle\!=\!0.94$ with a further increase in the electric field intensity. Additionally, among the experimental data, the maximal experimental polarization is $P_{\rm exp}^{\rm max}\!=\!0.097 \text{ C}/\text{m}^2\!=\!97 \text{ mC}/\text{m}^2$ \cite{furukawa1990electrostriction}. Hence, $\langle\sigma_z\rangle$ and $P_{\rm exp}$ are related as $\langle\sigma_z\rangle\!=\!(\langle\sigma_z^{\rm max}\rangle/P_{\rm exp}^{\rm max}) P_{\rm exp} \!\approx\! 9.7 P_{\rm exp}$. This relation can also be used to determine the remnant polarization $P^{\rm r}\!=\!54.8$ $\text{ mC}/\text{m}^2$, which corresponds to the experimental polarization as $E\!\to\! 0$ \cite{furukawa1990electrostriction}. Thus, we obtain the simulated remnant polarization at $E\!=\!0$ such that $\langle\sigma_z^{\rm r}\rangle\!=\!0.0548\!\times\!9.7\!=\!0.53$, and thus, $\lambda\!=\!0.345$ in the test simulations. Using the factor $P^{\rm r}/\langle\sigma_z^{\rm r}\rangle\!=\!54.8/0.53\!\approx\!103.4$, we define $P$ by $P\!=\!103.4 \langle\sigma_z\rangle$ to compare $\langle\sigma_z\rangle$ with $P_{\rm exp}$.

Next, we consider the relation between the external electric fields $E$ and $E_{\rm exp}$. First, \cite{proutorov2018finsler}
\begin{eqnarray}
\label{electrostriction}
\epsilon_0\Delta \epsilon E^2_{\rm exp}=\alpha E^2 \frac{k_BT}{a^3}
\end{eqnarray}
for the electrostriction energy, where $\epsilon_0\!=\!8.85\!\times\! 10^{-12}$ F/m is the vacuum permittivity, $\Delta \epsilon\!=\!0.25$ is the dielectric anisotropy, referenced from \cite{osaki1996high}, and $a$ is the lattice spacing. The parameter $\alpha$ is set to be relatively large ($\alpha\!=\!300$) because PVDF is a semicrystalline polymer, and thus, certain parts of the material are in the amorphous state. Moreover, owing to this reason, the PE loop becomes more rounded compared to that for the crystalline ferroelectrics.
According to the comparison of the dipole-field interactions, 
\begin{eqnarray}
\label{dipole-dipole}
P_{\rm exp}E_{\rm exp}=\langle\sigma_z\rangle Ef\frac{N}{V}\frac{k_BT}{a^3},
\end{eqnarray}
where $f$ is the number of monomers associated with one vertex, $N\!=\!10346$ is the total number of vertices, and $V\!=\!836$ is the cylinder volume (in units of $a^3$) at $E\!=\!0.15$. According to Eqs. (\ref{electrostriction}) and (\ref{dipole-dipole}), $E\!=\!\epsilon_0\Delta \epsilon (N\langle\sigma_z\rangle/\alpha V P_{\rm exp}) f E_{\rm exp}\!\simeq\!7.96\!\times\! 10^{-10}E_{\rm exp}$, using the abovementioned values, with $f\!=\!900$, where $E_{\rm exp}$ is in units of MV/m. Furthermore, $a\!\approx\!7.01\!\times\!10^{-9}$ m according to Eq. (\ref{electrostriction}), considering $k_BT\!=\!4\!\times\! 10^{-21}$. This $a$ is considerably larger than the van der Waals distance $10^{-10}$ m; i.e., $a$ lies in a reasonable range.

Moreover, the value $f\!=\!900$ is reasonable. This $f$ is slightly larger than the estimated real density $(N_A \rho V a^3)/(MN)\!\approx\!570$, where $\rho\!=\!1.97 \; {\rm g /(cm)^3}$ is the density of $\beta$-PVDF, $N_A$ is Avogadro's number, and $M$(=64 g/mol) is the molar mass of the PVDF monomer. This deviation occurs because the model is coarse grained, and each vertex corresponds to a lump of monomers; consequently, the monomer density is different from the actual density. Moreover, if the coefficient $\gamma$ of $\gamma S_1$ in Eq. (\ref{Hamiltonian}) is taken to be slightly larger than 1, the volume is expected to be slightly smaller than the current $V$; consequently, $f$ becomes smaller and is hence controllable.

%%%%%%%%%%%%%%%%%%%%%%%%%%%%%%%%%%%%%%%%%%%%%
\begin{figure}[h!]
\centering
\includegraphics[width=12.5cm]{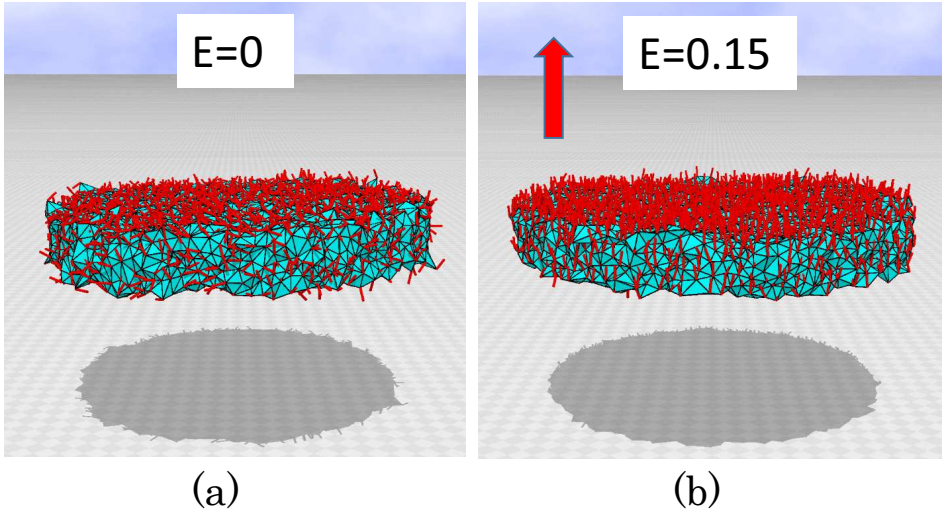}
\caption{\label{fig-2} (a) Snapshots of the lattice at (a) $E\!=\!0$ and (b) $E\!=\!0.15~(\Leftrightarrow E_{\rm exp}\!\simeq\!189 {\rm MV/m})$, where the scales of the graphics are the same. The small red cylinders on the surface denote the variable $\sigma$. The parameters are as follows: $\lambda\!=\!0.345$, $\kappa\!=\!1.8$, $\alpha\!=\!300$ and $v_0\!=\!0.1$. }
\end{figure}
%%%%%%%%%%%%%%%%%%%%%%%%%%%%%%%%%%%%%%%%%%%%%
The snapshots of the PVDF model are shown in Figs. \ref{fig-2}(a) and (b), where the external electric field is $E\!=\!0$ and $E\!=\!0.15$, corresponding to $E_{\rm exp}\!\simeq\!189$ MV/m. The small red cylinders on the surface denote the variable $\sigma$. $\sigma$ aligns along the $z$ axis even when $E\!=\!0$ because $\lambda$ is set as $\lambda\!=\!0.345$, corresponding to the ferroelectric behavior of the $\beta$ phase of the PVDF.

\begin{figure}[h!]
\centering
\includegraphics[width=12.5cm]{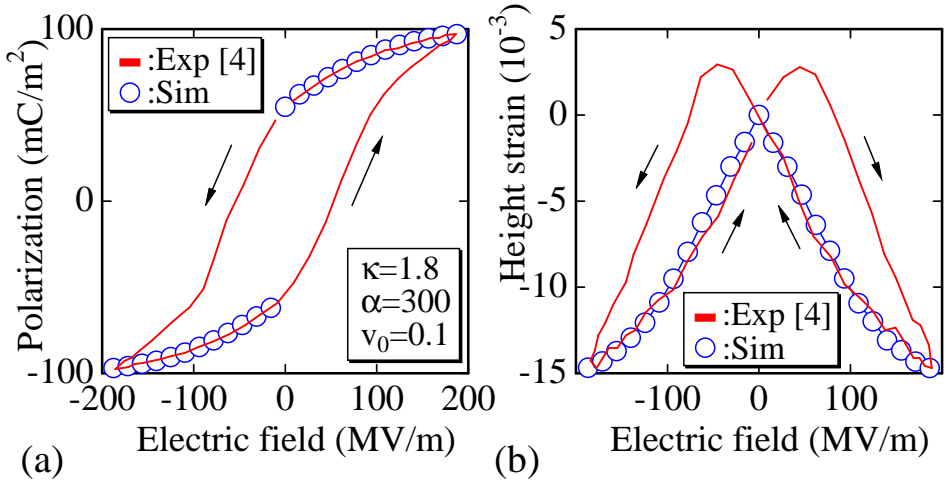}
\caption{\label{fig-3} (a) Polarizations $P_{\rm exp}$ and $P$ vs. electric field. (b) Height strain $\varepsilon_H$ vs. electric field. $\lambda\!=\!0.345$. }
\end{figure}

According to Fig. \ref{fig-3}(a), the simulation results of $P$ vs. $E$ are in agreement with $P_{\rm exp}$ vs. $E_{\rm exp}$. The influence of the other simulation parameters $\kappa$ and $v_0$, which are set as $\kappa\!=\!1.8$ and $v_0\!=\!0.1$, on the PE curve is almost negligible.

%%%%%%%%%%%%%%%%%%%%%%%%%%%%%%%%%%%%%%%%%%%%%
%%%%%%%%%%%%%%%%%%%%%%%%%%%%%%%%%%%%%%%%%%%%%

Fig. \ref{fig-3}(b) shows the relation between the height strain $\varepsilon_H$ and electric field. The simulated strain is obtained by measuring the mean height of the cylinder. The mean height is calculated from the vertices on the upper and lower surfaces at the central part inside a circle of radius $R_0\!=\!10\ell_0$, where $\ell_0$ is the mean bond length. We should note that the experimental loop has a butterfly shape, and the polymer shrinks (elongates) along the electric field when the field direction and polarization are parallel (antiparallel) to each other. The simulation results show that the height of the cylinder decreases with increasing electric field, in agreement with the experimentally observed data.

\begin{figure}[h]
\centering
\includegraphics[width=12.5cm]{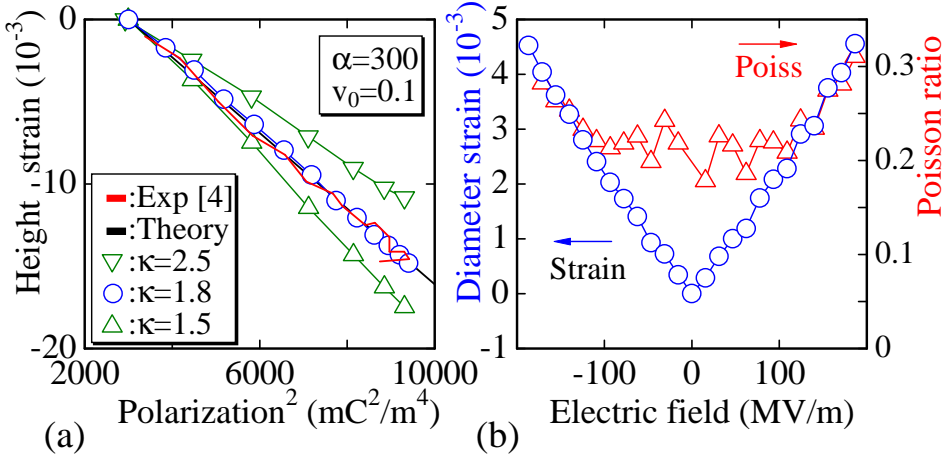}
\caption{\label{fig-4} (a) Height strain $\varepsilon_H$ vs. square of the polarization, with $\lambda\!=\!0.345$ and $\alpha\!=\!300$. The solid line (black) indicates the theoretical prediction with the electrostriction constant $\kappa_{33}\!=\!-2.4 \text{ m}^4/\text{C}^2$. (b) Diameter strain and Poisson ratio vs. electric field. The large fluctuation in the Poisson ratio is due to the numerical differentiation of the strain. }
\end{figure}

The dependence of the PVDF strain on the square of the polarization is linear and exhibits almost no hysteresis \cite{furukawa1990electrostriction}. This finding implies that the strain is directly related to the polarization rather than the electric field. Therefore, we expect that the electrostriction constant, i.e., the slope of the strain-polarization curve, can be controlled by $\kappa$. The solid line in Fig. \ref{fig-4}(a), denoted ``Theory'', is plotted using $P^2({S})\!=\!P^{2}(0)\!+\!(1/\kappa_{33}){S}$ with the electrostriction constant $\kappa_{33}\!=\!-2.4$ and experimental remnant polarization $P(0)$, where $S$ denotes the strain. The constant $|\kappa_{33}|$ decreases with increasing $\kappa$. The experimental sample involves the constant $\kappa_{33}\!=\!-2.4$, and this value is achieved in our model as $\kappa\!=\!1.8$.

%%%%%%%%%%%%%%%%%%%%%%%%%%%%%%%%%%%%%%%%%%%%%
%%%%%%%%%%%%%%%%%%%%%%%%%%%%%%%%%%%%%%%%%%%%%

The elongation of the cylinder diameter $D$ is calculated as $D\!=\!2\sqrt{V/(\pi H_0)}$ with the initial height $H_0$ at $E\!=\!0$. Using this $D$, we obtain the diameter strain $\varepsilon_D$, which is considered to be the ``nominal strain'' because $H_0$ is used to calculate $D$; hence, we estimate the Poisson ratio $\nu\!=\!\!-\varepsilon_D/\varepsilon_H$ (Fig. \ref{fig-4}(b)). $\nu$ tends to be close to 0.3 with increasing strain within the experimental electric field range. This value is comparable to the experimental value of 0.33 for $\beta$-PVDF obtained through mechanical testing \cite{vinogradov1999electro}.

Thus, we conclude that the FG model successfully reproduces the electromechanical properties of $\beta$-PVDF. Notably, the PE, strain-electric (SE) and Poisson ratio data are obtained with a single set of parameters.

%%%%%%%%%%%%%%%%%%%%%%%%%%%%%%%%%%%%%%%%%%%%%
\begin{figure}[h!]
\centering
\includegraphics[width=12.5cm]{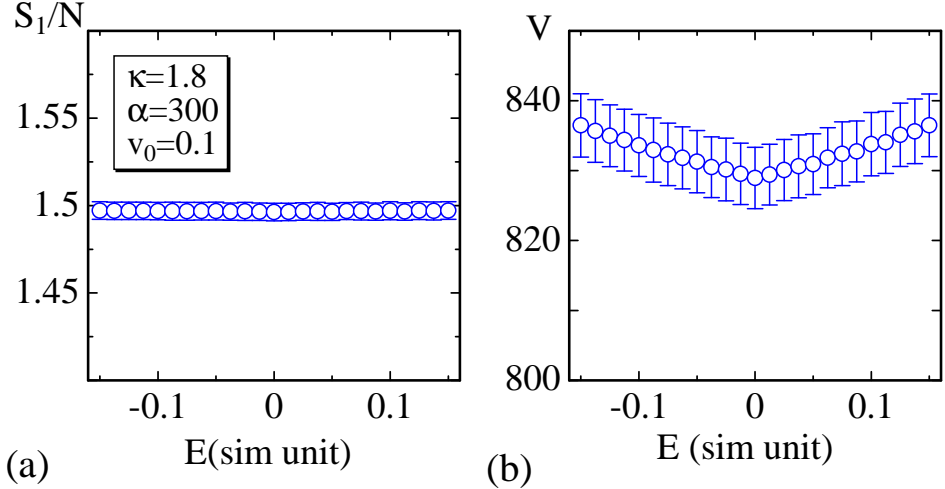}
\caption{\label{fig-5} (a) $S_1/N$ vs. $E$ and (b) volume $V$ vs. $E$, where the simulation unit is used for all the quantities.  }
\end{figure}
%%%%%%%%%%%%%%%%%%%%%%%%%%%%%%%%%%%%%%%%%%%%%
To ensure that the simulations are correctly performed, we plot $S_1/N$ vs. $E$, as shown in Fig. \ref{fig-5}(a), where the simulation unit is used for all the quantities including $E$.  As mentioned above, $S_1/N\!=\!1.5$ is satisfied. This finding implies that the constraints imposed on the lattice, such as the hard wall, do not influence the equilibrium property because this relation is satisfied only when no constraint is imposed on the vertex positions ${\vec r}_i$ \cite{koibuchi2014monte,takano2017j,proutorov2018finsler}. The relation of volume $V$ and $E$ in Fig. \ref{fig-5}(b) changes symmetrically with respect to the changes in $E$ and $-E$, as expected. 

We describe (i) the advantage of the FG model over the other models and (ii) how the FG model facilitates the understanding of such unusual and complex phenomena. To highlight these points, we intuitively express the basic idea. First, we must emphasize that  the elastic energy or bond potential $\gamma S_1\!=\!\gamma \sum_{ij}\Gamma_{ij} l_{ij}^2$ in Eq. (\ref{discrete-energy}) is of the form of the sum of  the ``effective coupling constant'' $\gamma\Gamma_{ij}$ and squared length $\ell_{ij}^2$ between molecules. The term ``molecules'' does not always express real molecules and is used in the context of coarse-grained particles, as described above.  

In the original model, in which the Euclidean metric is assumed, the effective coupling constant is  $\gamma\Gamma_{ij}\!=\!\gamma (\Leftrightarrow \Gamma_{ij}\!=\!1)$, and therefore, the elongation of the materials governed by this bond potential  must be isotropic. However, the elongation of the polymers under uniaxial forces, such as that of PVDF under electric fields, is anisotropic or direction dependent, and therefore, the former part, $\lambda$, of the bond potential should be direction dependent, because the latter part $\ell_{ij}^2$ is position dependent and does not depend on the polarization direction.

The question remains regarding the dependency of the constant $\lambda$ on the direction. One potential solution is to consider that the metric for the length unit of the local coordinates inside the material is dependent on the direction of the polarization vectors. In this manner, the deformation, such as that of PVDF, can be understood. The bond potential of the FG model is dependent on not only the squares of the distance but also the coupling constant, as described above. Moreover, the distance increases (decreases) for a small (large) coupling constant, as mentioned in Section \ref{model}, because the mean value of $S_1$  remains unchanged from that of the original model even if $S_1$ is deformed by the FG modeling prescription. Accordingly, the coupling constant ($\Leftrightarrow \Gamma_{ij}$) can be numerically or automatically determined by performing averaging over the scales of the molecular distance altered by fluctuations in the polarization direction.  

The key aspect is that the macroscopic property, such as the elongation of PVDF, varies depending on a certain position-dependent microscopic physical quantity of the material, such as the polarization vector of the PVDF. Depending on the value of the microscopic quantity, the unit Finsler length $v_{ij}$ is suitably defined, similar to that in Eq. (\ref{finsler-length}) with a suitable value of $v_0$,  and  the effective coupling constant ($\Leftrightarrow \Gamma_{ij}$) is automatically determined. In general, the macroscopic property obtained in this manner in the FG modeling is independent of the directional degree of freedom of the microscopic quantity and detailed information of the physical process such as the interactions of the electrons and atoms \cite{usui2016polymers}. Consequently, the FG modeling technique can be implemented easily in many models for complex phenomena and can be applied in other modeling techniques.

%%%%%%%%%%%%%%%%%%%%%%%%%%%%%%%
\section{Concluding remarks}
%%%%%%%%%%%%%%%%%%%%%%%%%%%%%%%
We demonstrate that the FG model can be applied to examine the electromechanical properties of ferroelectric polymers. The Monte Carlo simulation results are in agreement with the experimental data of $\beta$-PVDF. Both PE and SE field curves are reproduced using a single set of simulation parameters. Moreover, the simulated Poisson ratio is reasonable. 
The technique used for the PVDF is also applicable to ferroelectric ceramics such as ${\rm BaTiO_3}$. The magnetostriction of ferromagnetic materials will be examined in future work.

%\acknowledgments
The author H.K. thanks Laurent Chazeau for the valuable discussions on hysteresis curves. 
This work is supported in part by the Collaborative Research Project of the Institute of Fluid Science, Tohoku University, and by a JSPS Grant-in-Aid for Scientific Research on Innovative Areas "Discrete Geometric Analysis for Materials Design": Grant Number 20H04647 and JSPS KAKENHI Grant Number JP17K05149. The author V.E. thanks President Dr. Hiroshi Fukumura of Sendai KOSEN for the warm hospitality during the four-month stay from 2019 to 2020, which was supported in part by JSPS KAKENHI Grant Number JP17K05149.

\end{document}